\documentclass[10pt,conference]{IEEEtran}
\IEEEoverridecommandlockouts

\usepackage{amsmath,amssymb,amsfonts}
\usepackage{algorithmic}
\usepackage{graphicx}
\usepackage{textcomp}
\usepackage{xcolor}
\usepackage[colorlinks=true,urlcolor=teal,linkcolor=teal,citecolor=teal]{hyperref}
\usepackage{orcidlink}

\ifCLASSOPTIONcompsoc
  \usepackage[caption=false,font=normalsize,labelfont=sf,textfont=sf]{subfig}
\else
  \usepackage[caption=false,font=footnotesize]{subfig}
\fi

\usepackage[backend=bibtex,minbibnames=1, maxbibnames=3, style=ieee, mincitenames=1, maxcitenames=1, url=false]{biblatex}
\def\BibTeX{{\rm B\kern-.05em{\sc i\kern-.025em b}\kern-.08em
    T\kern-.1667em\lower.7ex\hbox{E}\kern-.125emX}}
    
\begin{document}

\title{From Classical Data 
to Quantum Advantage – Quantum Policy Evaluation on Quantum Hardware\\
\thanks{The project this report is based on was supported with funds from the German Federal Ministry for Economic Affairs and Climate Action in the funding program \emph{Quantum Computing – Applications for industry} under project number 01MQ22008A and 01MQ22008B. 
The sole responsibility for the report's contents lies with the authors.}
}

\author{\IEEEauthorblockN{Daniel Hein \orcidlink{0000-0002-8375-1592}$^{1,*}$, Simon Wiedemann \orcidlink{0000-0003-0473-3765}$^2$, Markus Baumann$^3$, Patrik Felbinger$^3$, Justin Klein$^3$, \\ Maximilian Schieder$^3$, Jonas Stein \orcidlink{0000-0001-5727-9151}$^{3}$, Daniëlle Schuman \orcidlink{0009-0000-0069-5517}$^{3}$, Thomas Cope$^{4}$, Steffen Udluft \orcidlink{0000-0002-5767-2591}$^{1}$
\thanks{*Corresponding author: hein.daniel@siemens.com}
\\}
\IEEEauthorblockA{
\small{
$^1$Siemens AG, $^2$Technical University of Munich, $^3$LMU Munich, $^4$IQM Germany
}}
}

\maketitle

\begin{abstract}
Quantum policy evaluation (QPE) is a reinforcement learning (RL) algorithm which is quadratically more efficient than an analogous classical Monte Carlo estimation.
It makes use of a direct quantum mechanical realization of a finite Markov decision process, in which the agent and 
the
environment are modeled by unitary operators and exchange states, actions, and rewards in superposition.
Previously, the quantum environment has been implemented and parametrized manually for an illustrative benchmark using a quantum simulator. 

In this paper, we demonstrate how these environment parameters can be learned from a batch of classical observational data through quantum machine learning (QML) on quantum hardware. 
The learned quantum environment is then applied in QPE to also compute policy evaluations on quantum hardware. 
Our experiments reveal that, despite challenges such as noise and short coherence times, the integration of QML and QPE shows promising potential for achieving quantum advantage in RL.
\end{abstract}


\section{Introduction}

Quantum reinforcement learning (QRL) is an emerging field of research at the intersection of quantum computing and reinforcement learning (RL), representing a promising new frontier for artificial intelligence (AI)~\cite{meyer2022survey}. 
As quantum computers leverage the unique principles of quantum mechanics, they present a paradigm shift in how computing tasks are approached, offering potential advantages over classical computing methods~\cite{nielsen2010quantum}. 
Although QRL algorithms with demonstrated theoretical advantages have been suggested~\cite{wang2021quantum, cherrat2023quantum,dunjko2016quantum,wiedemann2023quantum}, current quantum hardware limitations make their practical application often infeasible. 
Consequently, the focus has largely been on hybrid algorithms, wherein components such as the policy or 
the value function are augmented with variational quantum circuits (VQCs), albeit with heuristic methodologies that lack proof of inherent quantum advantage~\cite{bharti2022noisy,chen2020variational,eisenmannModelbasedOfflineQuantum2024}.

Quantum computers have shown potential to execute certain tasks significantly faster than classical machines, as evidenced by landmark algorithms in quantum search and factoring~\cite{grover1996fast,shor1999polynomial}. 
This raises the prospect that quantum computing might similarly enhance RL algorithms, potentially leading to more efficient learning processes. 
Existing theoretical and empirical data suggest that quantum interactions between an agent and its environment may reduce learning times.
By introducing a quantum policy iteration algorithm, the research presented in~\cite{wiedemann2023quantum} reveals a new source of quantum advantage. 
The method, known as quantum policy evaluation (QPE), achieves optimal policies with fewer agent-environment interactions and demonstrates enhanced sample efficiency compared to classical Monte Carlo methods.

QPE leverages a quantum mechanical realization of a finite Markov decision process (MDP). 
Within QPE, both the agent and environment are modeled using unitary operators, facilitating the exchange of states, actions, and rewards in superposition.

While in~\cite{wiedemann2023quantum}, the quantum environment has been implemented and parametrized manually for an illustrative benchmark using a quantum simulator, we demonstrate how these environment parameters can be learned from classical observational data through quantum machine learning (QML) on actual quantum hardware (Section~\ref{section:qml}).
In Section~\ref{section:qpe}, the learned quantum environment is then applied in QPE on quantum hardware to evaluate the performance of policies, also called policy values.
Our experiments reveal that, despite challenges such as noise and short coherence times, the integration of QML and QPE shows promising potential for achieving quantum advantage in RL.

\section{QPE on quantum hardware}

Current quantum RL techniques predominantly fall into two classes: quantum-enhanced agents operating in classical environments, and contexts where agent-environment interactions are quantum mechanical in nature. 
Research indicates that both strategies may provide substantial improvements over classical RL approaches. 

Specifically, methodologies for deep Q-learning and policy gradient models have been explored where classical neural networks are substituted with trainable variational quantum circuits, showcasing the quantum reinforcement learning potential~\cite{chen2020variational,lockwood2020reinforcement,moll2021comparing,skolik2022quantum,jerbi2021parametrized}.

Moreover, there is a line of works showing that, under specific assumptions on the RL problem at hand, quantum agent-environment interaction can yield agents that provably outperform \emph{any} classical agent in terms of rewards~\cite{dunjko2015framework}. 
For example, Grover-based learning strategies have been proposed for deterministic environments~\cite{dunjko2016quantum}, and amplitude estimation has been explored for stochastic settings~\cite{dunjko2017exponential,hamann2021quantum}. 
These approaches are related to QPE \cite{wiedemann2023quantum}, which plays a central role in this work.

\begin{figure}[t]
    \centering
    \includegraphics[width=1.0\columnwidth]{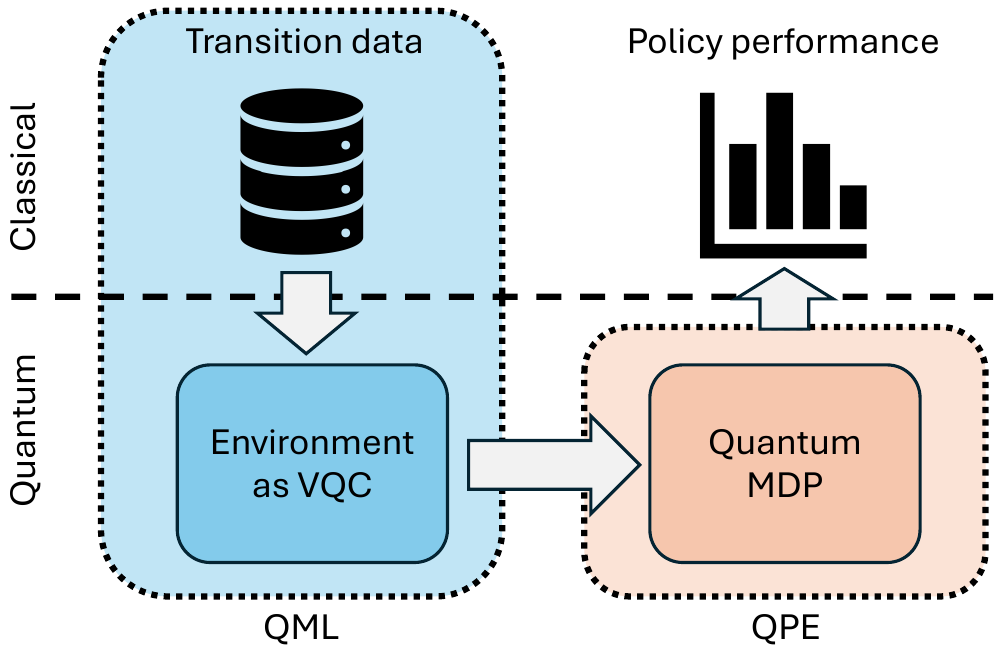}
    \caption{
    The schema depicts the context of this paper.
    The quantum-classical hybrid method QML is used to learn quantum gate parameters from classical RL transition data.
    The fully quantum QPE algorithm produces policy performance values.
    }
    \label{figure:schema}
\end{figure}

The QPE algorithm, as proposed by \citeauthor{wiedemann2023quantum}~\cite{wiedemann2023quantum}, takes a different approach from previous methods by directly searching for the optimal policy, rather than identifying rewarded actions to refine decision rules. 
Implementing QPE on a quantum computer requires the environment to be modeled using quantum circuits. 
How to obtain such a quantum model of the environment has not been studied so far. 
In this paper, we therefore focus on learning the parameters of quantum gates on classical data, so that the distribution obtained from quantum measurements aligns with the distribution of the batch of classical RL data. 
Quantum-classical hybrid QML is a well-established methodology that can automatically learn these gate parameters from classical data (see Figure~\ref{figure:schema}).
The experiments detailed in Section~\ref{section:qml} demonstrate that, for a two-armed bandit benchmark, quantum computer hardware from IBM can successfully learn circuit parameters.

In the subsequent step, the learned parameters can be utilized to parameterize the environment quantum circuit of the QPE. 
Figure~\ref{figure:schema} illustrates that, unlike quantum-classical hybrid QML, QPE is a purely quantum algorithm, with only the circuit measurements being transferred back to the classical domain after the circuit is executed. 

This paper is the first to demonstrate the application of QPE on actual quantum hardware. 
Section~\ref{section:qpe} presents our experimental findings using IonQ quantum hardware, showing that even with the current noisy conditions of today's hardware, meaningful results can be achieved for benchmarks of low complexity.

\section{Creating a quantum environment from classical data via quantum machine learning}
\label{section:qml}

In this section, we describe how to learn a quantum model of a two-armed bandit environment from classical data.

A two-armed bandit is a slot-machine-like RL environment consisting of a single state with two actions (arms), each yielding a reward of 0 or 1. As shown in~\cite{wiedemann2023quantum}, the interaction between an agent and a bandit can be modeled using a gate-based quantum circuit.

Our goal is to learn the quantum circuit corresponding to a fixed but unknown bandit from classical data. Before we explain our gradient-free learning process and discuss our experimental results, we briefly review the quantum model of agent-bandit interaction.

\subsection{VQCs for the two-armed bandit problem}

\begin{figure}[t]
    \centering
    \includegraphics[width=1.0\columnwidth]{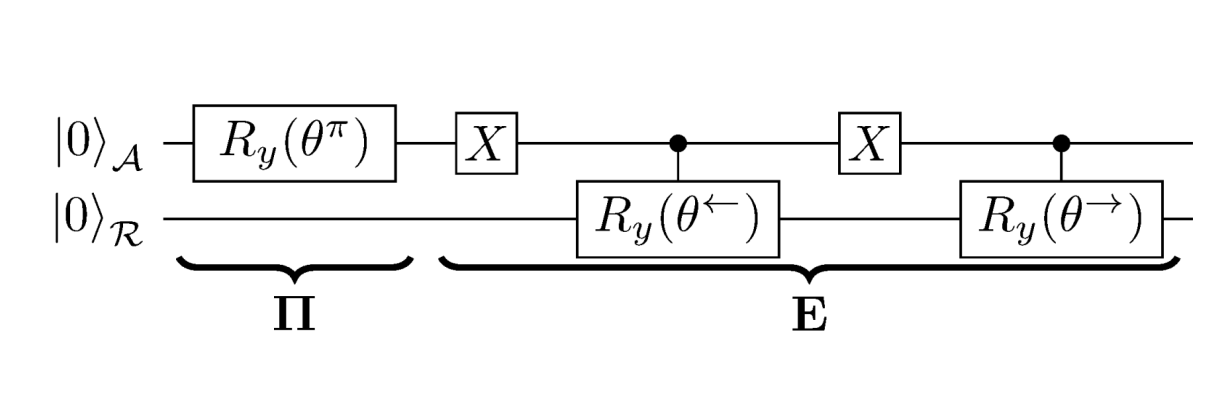}
    \caption{
    Parameterized quantum circuit for the two-armed bandit. 
    The Policy ($\Pi$) applies rotations to select actions, while the Environment ($E$) uses controlled-$R_y$ gates to generate reward probabilities. 
    Control qubits (solid circles) activate rotations based on the action state. 
    Used from~\cite{wiedemann2023quantum}.
    }
    \label{fig:circuit}
\end{figure}

The central idea is to encode both actions (pulling the left arm or the right arm) and their respective reward probabilities in a parameterized two-qubit circuit. 
By adjusting rotation parameters in this circuit, we can learn the probability of receiving a reward from each of the two arms from data, which are associated with specific qubit states. 
Our goal is to optimize these parameters.
The circuit, as shown in Figure~\ref{fig:circuit}, consists of two main segments: policy ($\Pi$) and environment ($E$). 
These segments implement rotation-based operations to explore and evaluate the reward probabilities for each arm.

When pulling one of the arms, a reward of either 0 or 1 is obtained. 
This corresponds to a Markov decision process with a single state and two possible actions, \textit{i.e.,} $\mathcal{A} = \{\leftarrow, \rightarrow\}$, where ``$\leftarrow$'' means ``pull the left arm'' and ``$\rightarrow$'' means ``pull the right arm''. 
The reward set is $\mathcal{R} = \{0, 1\}$

The Markov decision process dynamics are determined by the two probabilities $p(0 \mid \leftarrow)$ and $p(0 \mid \rightarrow)$ of losing when pulling the left or right arm.

The actions are encoded via 
\begin{equation}
    \left|\leftarrow\right\rangle = \left|0\right\rangle_{\mathcal{A}}, \quad \left|\rightarrow\right\rangle = \left|1\right\rangle_{\mathcal{A}},
\end{equation}
which can be done by using a single qubit. 
Another qubit is used to encode the rewards as 
\begin{equation}
    \left|0\right\rangle = \left|0\right\rangle_{\mathcal{R}}, \quad \left|1\right\rangle = \left|1\right\rangle_{\mathcal{R}}.
\end{equation}

\begin{figure} 
    \centering
  \subfloat[Left arm ($\leftarrow$) quantum circuit.\label{fig:circuit-left}]{%
       \includegraphics[width=0.485\textwidth]{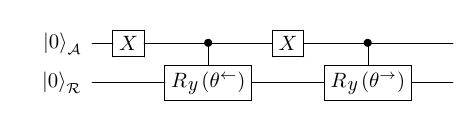}}
    \\
  \subfloat[Right arm ($\rightarrow$) quantum circuit.\label{fig:circuit-right}]{%
        \includegraphics[width=0.485\textwidth]{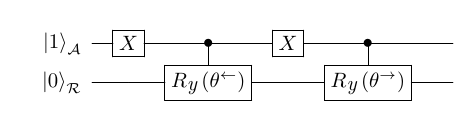}}
  \caption{
    (a) The action qubit $|0\rangle_{\mathcal{A}}$ is flipped to $|1\rangle_{\mathcal{A}}$ via an $X$-gate, activating a controlled-$\mathcal{R}_y(\theta^{\leftarrow})$ rotation on the reward qubit. 
    A final $X$-gate resets the action qubit. 
    The reward probability is governed by $\theta^{\leftarrow}$.
    (b) The action qubit $|1\rangle_{\mathcal{A}}$ is temporarily flipped to $|0\rangle_{\mathcal{A}}$, deactivating the first controlled rotation. 
    After restoring $|1\rangle_{\mathcal{A}}$, a controlled-$\mathcal{R}_y(\theta^{\rightarrow})$ rotation sets the reward probability via $\theta^{\rightarrow}$.
    }
  \label{fig:circuit-left-right} 
\end{figure}

The gates used in the quantum circuit are defined as follows:
\begin{itemize}
    \item The $X$-gate, responsible for flipping the action qubit, has the matrix representation:
    \begin{equation}
        X = \begin{bmatrix}
        0 & 1 \\
        1 & 0
        \end{bmatrix}.
    \end{equation}
    \item The $R_y(\theta)$-gate, which performs a rotation of the qubit around the $y$-axis by an angle $\theta$, is represented as:
    \begin{equation}
        R_y(\theta) = \begin{bmatrix}
        \cos\left(\frac{\theta}{2}\right) & -\sin\left(\frac{\theta}{2}\right) \\
        \sin\left(\frac{\theta}{2}\right) & \cos\left(\frac{\theta}{2}\right)
        \end{bmatrix}.
    \end{equation}
\end{itemize}
For our specific problem, the optimization of the rotation parameters $\theta^{\leftarrow}$ and $\theta^{\rightarrow}$ in the environment is of primary importance. 
To simplify the calculations and focus on this optimization, we fix the policy outcome and separately analyze the circuit when the policy selects either $\left|\leftarrow\right\rangle = \left|0\right\rangle_{\mathcal{A}}$ or $\left|\rightarrow\right\rangle = \left|1\right\rangle_{\mathcal{A}}$.
The corresponding quantum circuits for these actions are depicted in Figure~\ref{fig:circuit-left-right}.

For the left arm, the action qubit starts in the state $|0\rangle_{\mathcal{A}}$. 
An $X$-gate is applied, flipping the action qubit to $|1\rangle_{\mathcal{A}}$. 
Then, a controlled rotation $C\mathcal{R}_y(\theta^{\leftarrow})$ is applied, where the rotation angle $\theta^{\leftarrow}$ is conditioned on the state of the action qubit. 
Finally, a second $X$-gate is applied to return the action qubit to its original state. 
The final state for the left arm becomes:
\begin{equation}
    |\psi^{\leftarrow}\rangle = \cos\left(\frac{\theta^{\leftarrow}}{2}\right) |00\rangle + \sin\left(\frac{\theta^{\leftarrow}}{2}\right) |01\rangle.
\end{equation}

Here, the state $|00\rangle$ corresponds to no reward ($\mathcal{R} = |0\rangle_{\mathcal{R}}$), and the state $|01\rangle$ corresponds to a reward ($\mathcal{R} = |1\rangle_{\mathcal{R}}$). 
The probability of receiving a reward (measuring $\mathcal{R} = |1\rangle$) for the left arm is:
\begin{equation}
    P^{\leftarrow}(\mathcal{R} = 1) = \sin^2\left(\frac{\theta^{\leftarrow}}{2}\right).
\end{equation}

For the right arm, the action qubit starts in the state $|1\rangle_{\mathcal{A}}$ which yields the probability of receiving a reward (measuring $\mathcal{R} = |1\rangle$) for the right arm:
\begin{equation}
    P^{\rightarrow}(\mathcal{R} = 1) = \sin^2\left(\frac{\theta^{\rightarrow}}{2}\right).
\end{equation}

\subsubsection*{Application of Results}

Building on this analysis, and with the winning frequencies derived from the data, the rotation angles $\theta^{\leftarrow}$ and $\theta^{\rightarrow}$ can be directly calculated using the inverted following formulas:
\begin{equation}
    \theta^{\leftarrow} = 2 \arcsin \left( \sqrt{f_{\text{data}}^\leftarrow} \right),
\end{equation}
\begin{equation}
    \theta^{\rightarrow} = 2 \arcsin \left( \sqrt{f_{\text{data}}^\rightarrow} \right),
\end{equation}
where $f_{\text{data, left}}$ and $f_{\text{data, right}}$ denote the measured winning frequencies for the left arm and right arm, respectively.

However, while these formulas offer a direct theoretical method for determining the optimal rotation parameters from data, this approach circumvents the core objective of inferring these parameters within a realistically noisy quantum environment. 
Consequently, the theoretically derived parameters serve merely as reference points in our optimization process. 
In practice, our goal is to learn the rotation parameters $\theta^{\leftarrow}$ and $\theta^{\rightarrow}$ directly from data through a data-driven learning framework.

\subsection{Learning circuit parameters from classical data}
\label{subsec:learning}

In a two-armed bandit setting, we consider two options, labeled ``left'' $(\leftarrow)$ and ``right'' $(\rightarrow)$.
Each arm has an empirically estimated reward relative frequency, $f^{\leftarrow}_{\text{data}}$ and $f^{\rightarrow}_{\text{data}}$, derived from classical data.

Unlike purely noise-free simulations, our experiments are performed on an existing quantum device, subject to real hardware constraints such as gate errors, decoherence, and measurement noise. 
For each arm and a given parameter set $\boldsymbol{\theta} = (\theta^{\leftarrow}, \theta^{\rightarrow})$, we execute the circuit $M$ times (\textit{i.e.,} perform $M$ shots) on the quantum hardware. 
Let $N_R^\square(\boldsymbol{\theta})$ be the number of shots that return a ``reward'' outcome for arm $\square \in \{\leftarrow, \rightarrow\}$ under the current parameters. 
The measured relative frequency of success for arm $\square$ is $f^\square_{\text{meas}}(\boldsymbol{\theta}) = N_R^\square(\boldsymbol{\theta})/{M}$.
We aim to determine the parameters $\boldsymbol{\theta}$ that align the measured relative frequencies $f\square_{\text{meas}}(\boldsymbol{\theta})$ as closely as possible with the classical relative frequencies $f^\square_{\text{data}}$.

\subsection{Gradient-free optimization}
\label{subsec:gradient-free}

Gradient-free methods are particularly suited to noisy quantum hardware due to their robustness against stochastic perturbations in the cost function evaluation \cite{Franken2022evolution,Singh2023optimizers}. 
These methods explore the parameter space iteratively without requiring explicit gradients, updating the parameters $\boldsymbol{\theta}^{(k)}$ as follows:
\begin{equation}
    \boldsymbol{\theta}^{(k+1)} = \boldsymbol{\theta}^{(k)} + \Delta\boldsymbol{\theta},
    \label{gradient-free-formula}
\end{equation}
where $\Delta\boldsymbol{\theta}$ is determined by the optimization routine based on the measured outcomes.
The mean squared error (MSE) between the predicted relative frequencies, $f_{\text{meas}}$, and the observed frequencies, $f_{\text{data}}$ is used as cost function.
For the experimental optimization approach, the process unfolds as:
\begin{enumerate}
    \item \textbf{Frequency calculation:} 
    Compute the empirical relative frequencies $f_{\text{data,left}}$ and $f_{\text{data,right}}$ from the given data.

    \item \textbf{Quantum circuit execution:} 
    Choose a current set of parameters $\boldsymbol{\theta}$ and execute the corresponding quantum circuit with $8000$ shots.

    \item \textbf{Relative frequency measurement:} 
    From the measurement outcomes, estimate the relative frequencies $f_{\text{meas,left}}(\boldsymbol{\theta})$ and $f_{\text{meas,right}}(\boldsymbol{\theta})$.

    \item \textbf{Loss evaluation:}
    Compute the 
    MSE to assess the alignment between $f_{\text{data}}$ and $f_{\text{meas}}$.

    \item \textbf{Parameter update:}
    Update $\boldsymbol{\theta}$ to minimize the MSE.
    \item \textbf{Repeat:}
    Iterate steps 2--5 for a predefined number.
\end{enumerate}

\subsection{Experiments}

To optimize the rotational parameters $\theta^\leftarrow$ and  $\theta^\rightarrow$, we employed gradient-free optimization techniques available in the \texttt{scipy.optimize.minimize} library\footnote{\url{https://scipy.org}}. 
This library offers a variety of algorithms, including gradient-free methods that are particularly well-suited for noisy quantum hardware. 
Among these, we utilized COBYLA (constrained optimization by linear approximations) for two different parameter settings. 
COBYLA operates by approximating the optimization problem using linear constraints and iteratively refining the solution. 
It is robust to noise in cost function evaluations and does not require explicit gradient information~\cite{powell1994direct}.

While COBYLA provides a robust framework for parameter optimization, the noise inherent in quantum hardware requires noise mitigation strategies to ensure accurate and reliable results. 
For this purpose, we integrated the FireOpal\footnote{\url{https://q-ctrl.com/fire-opal}} package from Q-CTRL into our optimization workflow.

\subsection{Q-CTRL Noise Mitigation}

Q-CTRL is a company that is focusing on quantum noise mitigation and has developed an automated workflow for deterministic error suppression in quantum algorithms. 
Their software package, FireOpal, has been experimentally demonstrated on commercial hardware from IBM, showing improvements of over 1000× compared to the best expert-configured implementations using tools available on these platforms~\cite{qctrl_paper}. 

In this work, the \textit{iterate} function from Q-CTRL was utilized to integrate Q-CTRL's noise mitigation capabilities into the gradient-optimization process for the parameters of our two-armed bandit problem.

\subsection{Results}
\label{subsec:results_learning}

The achieved results for the optimized parameters are displayed in Table~\ref{tab:optimizer_results_gradient_} rounded to two decimal digits.

\begin{figure*}[ht]
    \centering
  \subfloat[70\% and 20\% learning curve\label{fig:exp1-learningcurve}]{
       \includegraphics[width=0.45\linewidth]{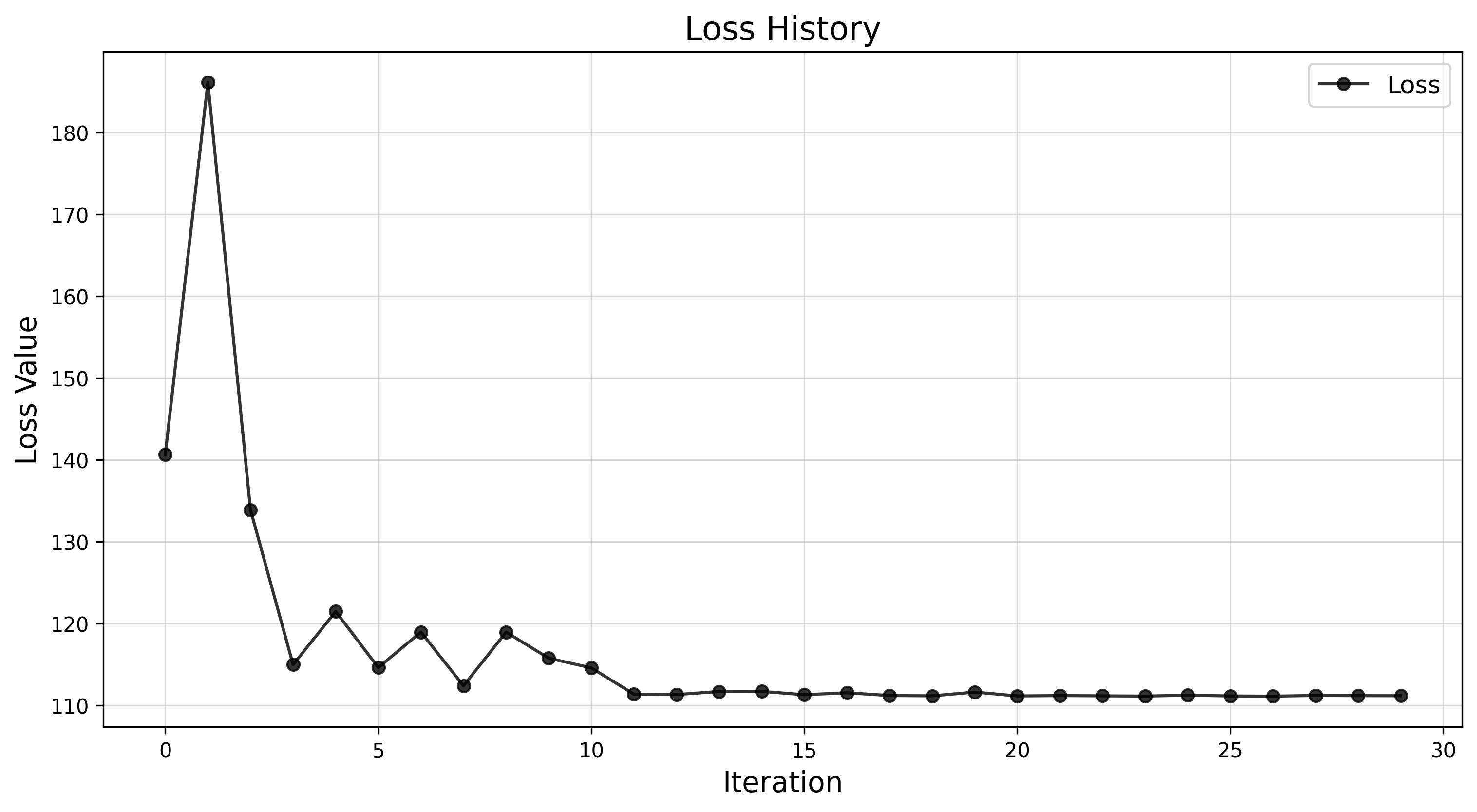}}
    \hfill
  \subfloat[70\% and 20\% learned parameters\label{fig:exp1-parameters}]{
        \includegraphics[width=0.45\linewidth]{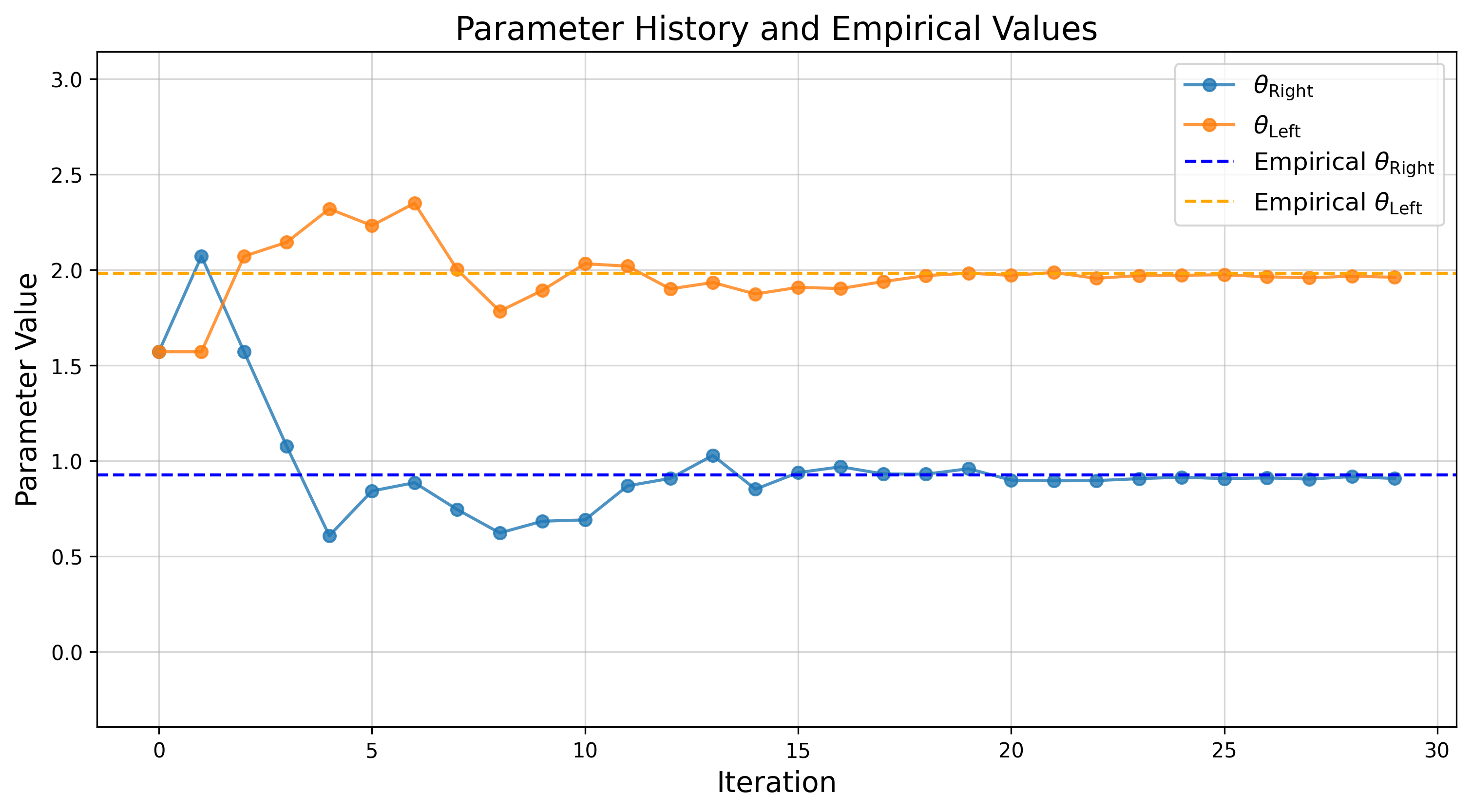}}
    \\
  \subfloat[0\% and 50\% learning curve\label{fig:exp2-learningcurve}]{
        \includegraphics[width=0.45\linewidth]{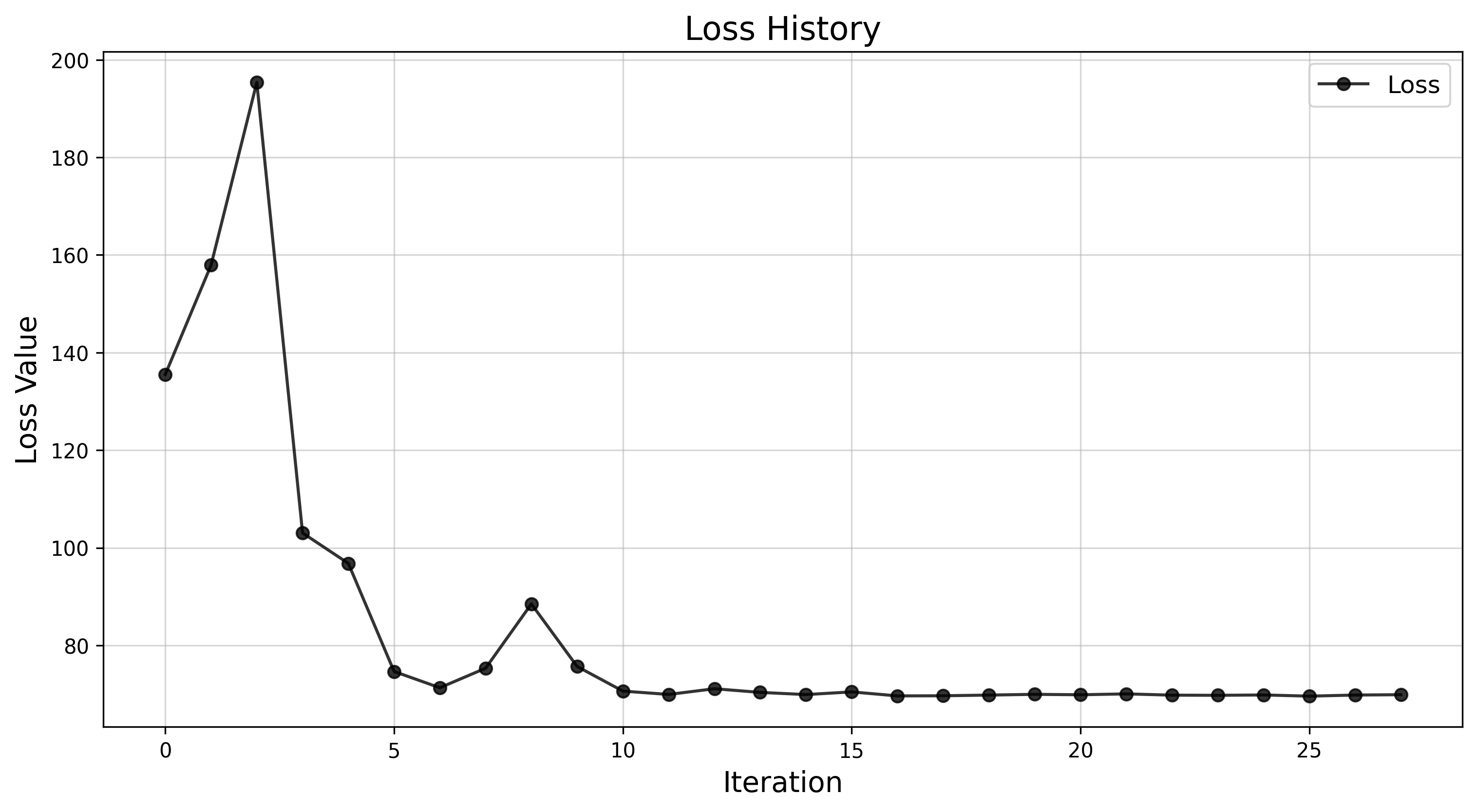}}
    \hfill
  \subfloat[0\% and 50\% learned parameters\label{fig:exp2-parameters}]{
        \includegraphics[width=0.45\linewidth]{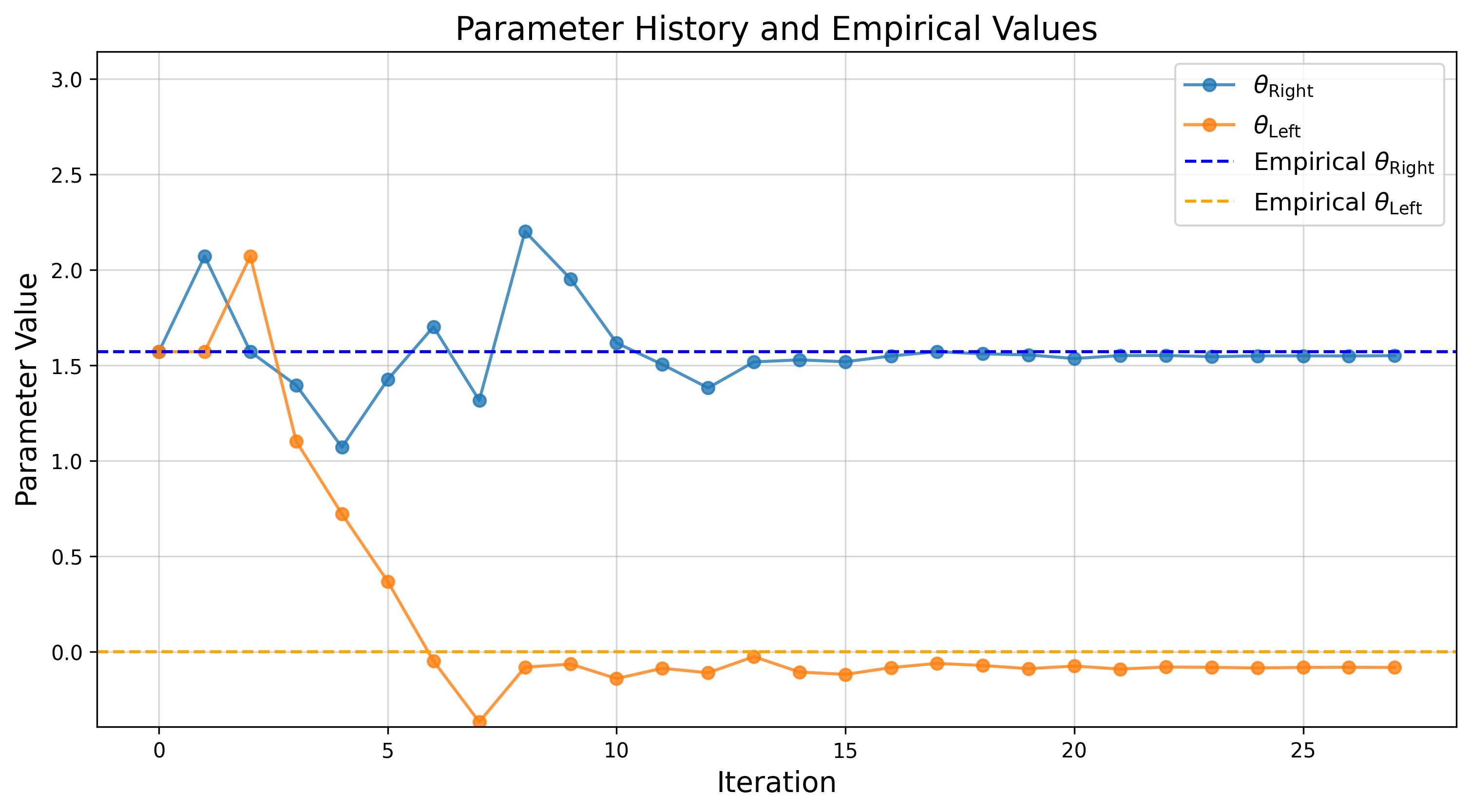}}
  \caption{Comparison of the theta parameter evolution for experiments with winning probabilities of 70\%/20\% and 0\%/50\%.}
  \label{fig:all-experiments} 
\end{figure*}

With noise mitigation through Q-CTRL the deviation is 0.015 for both parameters, as shown in Figure~\ref{fig:exp1-parameters}.
The optimal loss within the parameter setting corresponding to a winning probability of 70\% and 20\% is $\mathcal{L}_{opt} \approx 111.3$ (see Figure~\ref{fig:exp1-learningcurve}).

Figure~\ref{fig:exp2-parameters} illustrates the optimization process for a winning probability of 0\% and 50\%, highlighting the performance at the edges of the parameter space. 
The loss curve for this experiment is displayed in Figure~\ref{fig:exp2-learningcurve} with the optimal loss being $\mathcal{L}_{opt} \approx 69.3$.

The results with the gradient-free optimization approach underline the effectiveness of gradient-free methods in optimizing parameters within this QML framework.

\begin{table}[t]
    \caption{Results for two winning probabilities (70\%/20\% and 0\%/50\%)}
    \centering
    \begin{tabular}{|rcl|c|c|c|c|}
        \hline
        \multicolumn{3}{|c|}{\textbf{Winning prob.}} & \textbf{Final}& \textbf{Final} & \textbf{Empirical} & \textbf{Empirical}  \\
        &&& $\theta^\leftarrow$ & $\theta^\rightarrow$ & $\theta^{\leftarrow}_{emp}$ & $\theta^{\rightarrow}_{emp}$ \\
        \hline
        70\% \hspace{-1.3em} & and & \hspace{-1em}20\% & 1.96 & 0.91 & 1.98 & 0.93 \\
        0\% \hspace{-1.3em} & and & \hspace{-1em}50\% & -0.08 & 1.55 & 0 & 1.57 \\
        \hline
    \end{tabular}
    \label{tab:optimizer_results_gradient_}
\end{table}

\section{Performing QPE on IonQ hardware}
\label{section:qpe}

\begin{figure*}[ht]
    \centering
\includegraphics[width=0.85\textwidth]{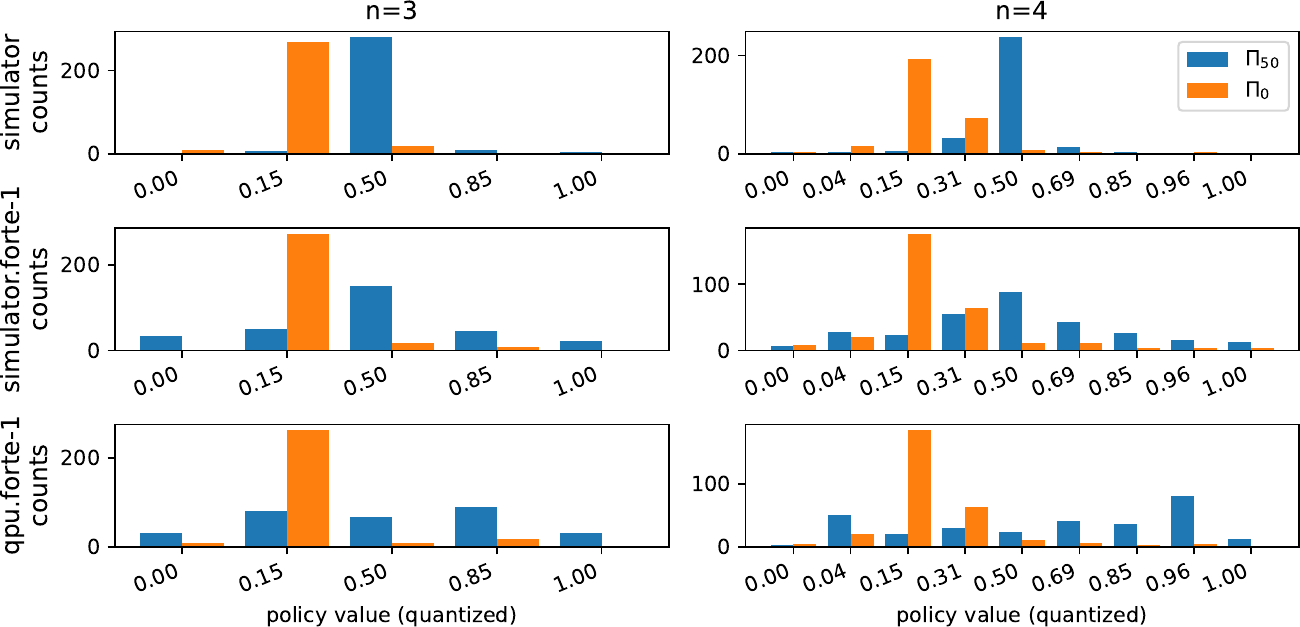}
    \caption{
    Empirical output distributions over policy values $\tilde{v}$ estimated with QPE for the two-armed bandit for different values of $n$ on an ideal simulator (top row), a noisy simulator (center row), and the real \texttt{forte-1} quantum computer (bottom row).
    Each bar represents the number of times we observed a specific estimate $\tilde{v}$ of the true policy value $v$.
    }
    \label{fig:bandit_qpe}
\end{figure*}

Here, we present experiments on the QPE algorithm~\cite{wiedemann2023quantum}, which we use to estimate the value of a policy $\Pi$ on a two-armed bandit environment $E$.
Our goal is to study how noise and limited coherence times affect the performance of QPE on real quantum hardware.

\subsection{Quantum policy evaluation}

Before we describe our experiments, we briefly summarize the QPE algorithm and its properties. For details, we refer to the original work by \citeauthor{wiedemann2023quantum}~\cite{wiedemann2023quantum}.

QPE uses \emph{quantum phase estimation} to estimate the value $v$ of a policy $\Pi$ in an environment $E$ given quantum access to them as described above. 
The complexity of QPE can be measured in terms of the number of \emph{quantum samples} (qsamples) from quantum models of the policy $\Pi$ and the environment $E$ needed to approximate the value of a policy up to a user-specified error. 
The qsample complexity of QPE is quadratically lower than that of the best possible analogous classical Monte Carlo approach. 

We now restate the approximation error of QPE.
For simplicity, we assume the value $v$ of the policy $\Pi$ satisfies $v \in (0,1)$. 
QPE is a stochastic algorithm which produces an estimate $\tilde{v}$ of the true value $v$ that obeys%
\begin{equation}
    \label{eq:qpe_bound}
    P\big(|\tilde{v} - v| \leq \epsilon(n) \big) \geq \frac{8}{\pi^2}.
\end{equation}
Here $\epsilon(n) > 0$ is an upper bound on the approximation error that satisfies $\epsilon(n) \in \mathcal{O}(1/2^n)$. The hyperparameter $n \in \mathbb{N}$ is specified by the user in order to achieve a desired approximation accuracy.

QPE uses $\mathcal{O}(1/\epsilon(n))$ qsamples of the quantum policy $\Pi$ and environment $E$. 
Higher values of $n$ yield a lower approximation error which comes at the cost of more qsamples.

Finally, note that, when implemented on a gate-based quantum computer with qubits, the output $\tilde{v}$ of QPE is quantized: for a given $n$, QPE uses $n$ qubits to represent the approximation $\tilde{v}$ of the true value $v$
and returns one of $2^{(n-1)}+1$ different values for $\tilde{v}$.

\subsection{Experimental setup}

We consider a two-armed bandit environment $E$ whose left/right arm yields a return of $1$ with probabilities $70\%$ and $20\%$, as learned in Section~\ref{section:qml}. Our goal is to estimate the value of two  separate policies $\Pi_{50}$ and $\Pi_{0}$ that choose the left arm with probability $\Pi_{50}(\leftarrow) =50\%$ and $\Pi_{0}(\leftarrow) =0\%$, respectively. The true values of these policies are $v_{50} = 0.45$ and $v_{0}=0.2$.

We implemented QPE on \emph{IonQ}'s \texttt{forte-1} quantum computer which is gate-based and uses trapped ions~\cite{chen2024benchmarking}.
In our implementation, we followed the approach outlined by \citeauthor{wiedemann2023quantum}~\cite{wiedemann2023quantum}, who describe a gate-based implementation of QPE for two-armed bandits.

As larger values for $n$ result in lower approximation error (see Equation \eqref{eq:qpe_bound}) but higher gate complexity, and, therefore, more possibility for error accumulation, we tried $n=3$ and $n=4$, which resulted in QPE circuits with 5 and 6 qubits consisting of 412 and 883 one- and two-qubit gates, respectively.
We ran each circuit for $300$ times to produce an empirical estimate of the measurement distribution.

\subsection{Experimental results}

The results of our experiments are summarized in Figure~\ref{fig:bandit_qpe}. We first discuss the results for policy $\Pi_{50}$ (blue bars). When using a noiseless simulator (top row), QPE returns a close approximation $\tilde{v}$ of the value $v=0.45$ with high probability. As predicted by the theory, the approximation improves for larger $n$.
In the ideal setup, QPE does not suffer from the increased circuit depth that is attached to larger $n$. 

When we switch to IonQ's noisy simulator that has been designed to mimic the real \texttt{forte-1} device (center row of Figure~\ref{fig:bandit_qpe}), we observe that deep circuits for large $n$ harm QPE's performance, as more shots are farther away from the true value $v=0.45$ than in the ideal case. As expected, the effect becomes more severe as $n$ grows.

As can be seen in the bottom row of Figure~\ref{fig:bandit_qpe}, the noise of the real \texttt{forte-1} device affects QPE's performance stronger than the simulated noise. 

For policy $\Pi_{0}$ (orange bars), neither simulated nor real noise have noticable negative effects on QPE. The results obtained with the noiseless simulator, the noisy simulator and the real hardware are almost identical. This is likely because for the deterministic $\Pi_0$ policy,  the environment effectively collapses to the single $R_y(\theta^{\leftarrow})$ gate (see Figure \ref{fig:circuit}), thereby reducing potential sources of error.

\section{Conclusion}

In this work, we have integrated two quantum computing approaches, QML and QPE, to demonstrate the feasibility of QRL on actual quantum hardware. 
QML is employed to automatically learn quantum environment parameters from classical offline RL batch data. 
These parameters are then used by QPE to determine policy values. 
Our experiments indicate that, for the two-armed bandit benchmark, it is possible to learn circuit parameters that result in satisfactory model performance on real quantum hardware. 
However, only the most basic circuit configuration in QPE produced the expected outcomes. 
Challenges such as noise and errors during gate operations complicate the application of this QRL approach on current quantum hardware. 
Nonetheless, the continuous advancements in hardware 
as well as in error mitigation techniques
present a promising future for QRL methods, with potential quantum advantage.

\printbibliography
 
\end{document}